 \documentclass[journal=jpcl,layout=twocolumn]{achemso}
\usepackage{color}
\usepackage{graphicx}
\usepackage{amsmath}
\usepackage{amsfonts}
\usepackage[utf8]{inputenc}      
\usepackage{mathtools}  
\usepackage{braket}              
\usepackage{bm}
\usepackage{units}
\usepackage[usenames,dvipsnames,svgnames,table]{xcolor}

\newcommand{\pa}{\nabla_1}

\newcommand{\prv}{\nabla_{\vec{r}}}

\newcommand{\vp}{\varphi}
\newcommand{\ve}{\varepsilon}

\renewcommand{\vec}{\bm}

\author{Jakub Koc\'ak}
\affiliation{Laboratorium für Physikalische Chemie, ETH Z\"urich, Vladimir-Prelog-Weg 2, 8093 Z\"urich, Switzerland}

\author{Eli Kraisler}
\affiliation{Fritz Haber Center for Molecular Dynamics, Institute of Chemistry, The Hebrew University of Jerusalem, 91904 Jerusalem, Israel}

\author{Axel Schild}
\email{axel.schild@phys.chem.ethz.ch}
\affiliation{Laboratorium für Physikalische Chemie, ETH Z\"urich, Vladimir-Prelog-Weg 2, 8093 Z\"urich, Switzerland} 

\title{Charge-transfer steps in Density Functional Theory from the perspective of the Exact Electron Factorization}


\begin{document}

  
  \begin{abstract}
    When a molecule dissociates, the exact Kohn-Sham (KS) and Pauli potentials may form step structures.
    Reproducing these steps correctly is central for the description of dissociation and charge-transfer processes in density functional theory (DFT):
    The steps align the KS eigenvalues of the dissociating subsystems relative to each other and determine where electrons localize.
    While the step height can be calculated from the asymptotic behavior of the KS orbitals, this provides limited insight into what causes the steps.
    We give an explanation of the steps with an exact mapping of the many-electron problem to a one-electron problem, the exact electron factorizaton (EEF).
    The potentials appearing in the EEF have a clear physical meaning that translates to the DFT potentials by replacing the interacting many-electron system with the KS system.
    With a simple model of a diatomic, we illustrate that the steps are a consequence of spatial electron entanglement and are the result of a charge transfer.
    From this mechanism, the step height can immediately be deduced.
    Moreover, two methods to approximately reproduce the potentials during dissociation are proposed.
    One is based on the states of the dissociated system, while the other one is based on an analogy to the Born-Oppenheimer treatment of a molecule.
    The latter method also shows that the steps connect adiabatic potential energy surfaces.
    The view of DFT from the EEF thus provides a better understanding of how many-electron effects are encoded in a one-electron theory and how they can be modeled.
  \end{abstract}
  
  \maketitle
  
  \vspace{10mm}
  
  The many-electron problem is central for the prediction of molecular structures, reaction mechanisms, and material properties \cite{dirac1929}.
  Density functional theory (DFT) \cite{hohenberg1964} is an elegant approach to solve the many-electron problem by reducing it to a one-electron problem \cite{dreizler1990,ullrich2013}.
  Its most common variant, Kohn-Sham (KS) DFT \cite{kohn1965}, has been successfully used in a variety of cases where the nuclei are close to their equilibrium configuration.
  However, existing functionals used in time-independent and time-dependent KS-DFT \cite{runge1984} typically have problems to describe dissociation of a molecule and charge-transfer processes \cite{hellgren2012a,maitra2016,maitra2017}.
  Although some promising development have been made very recently \cite{chen2017,su2020,mei2020}, it is, in general, a major limitation: 
  Most chemical reactions involve dissociation and charge transfer, and thus are only partly amenable to applied KS-DFT at the moment.
  
  It was found previously that steps can appear in the exact KS potential $v^{\rm KS}$ during dissociation and charge transfer (see, e.g., \cite{gritsenko1996b,elliott2012,kohut2016,hodgson2016,hodgson2017} and citations therein), that they are related to the derivative distcontinuity and the problem of describing charge-transfer in DFT \cite{tozer2003}, and that many density functionals do not describe these steps correctly \cite{kraisler2020b}.
  Steps were also found in the Pauli potential $v^{\rm P}$ \cite{kraisler2020} that appears in another variant of DFT, orbital-free (OF) DFT \cite{levy1984,march1986,march1987,levy1988}.
  The steps can be rationalized with the asymptotic behavior of the KS orbitals, as these are the eigenfunctions of $v^{\rm KS}$ and provide the one-electron density that is the ground state of $v^{\rm KS}+v^{\rm P}$.
  Notwithstanding this, the asymptotic behavior of the KS orbitals provides little understanding of the physical reason for the steps.
  
  In this Letter, we show how and why steps form with the help of quantities derived from the interacting many-electron system, and we relate them to the KS system. 
  For this purpose, we use the exact electron factorization (EEF) \cite{schild2017,kocak2020}.
  The EEF is an exact theory which reduces an $N$-electron system to a one-electron system by expressing the wavefunction $\psi$ of the electrons as a product of a marginal and a conditional wavefunction \cite{hunter1986,buijse1989,gritsenko1996b,abedi2010,abedi2012,gidopoulos2014,schild2017,kocak2020},
  \begin{align}
    \psi(1,\dots,N) = \chi(1) \phi(2,\dots,N;1)
    \label{eq:eef}
  \end{align}
  with the partial normalization condition 
  \begin{align}
    \Braket{\phi(2,\dots,N;1)|\phi(2,\dots,N;1)}_{2,\dots,N} \stackrel{!}{=} 1,
    \label{eq:pnc}
  \end{align}
  valid for all values of the electronic coordinate $\vec{r}_1$.
  $|\phi(2,\dots,N;1)|^2$ is the conditional probability density of finding $N-1$ electrons at $\vec{r}_2, \dots, \vec{r}_N$, given one electron is at $\vec{r}_1$.
  We call those $N-1$ electrons the conditioned system.
  The bra-ket $\Braket{\dots|\dots}_{2,\dots,N}$ represents the complex-valued scalar product with integration over coordinates $\vec{r}_2, \dots, \vec{r}_N$.
  As arguments of a function, we use either the coordinates $\vec{r}_j \in \mathbb{R}^3$ or the number $j$.
  For simplicity we do not consider spin here; the extension to include spin is straightforward but needs additional notation which would complicate the discussion.
  
  We assume $\psi$ is given by the Schr\"odinger equation
  \begin{align}
    \left( -\sum_{j=1}^N \frac{\nabla_j^2}{2} + V(1,\dots,N) \right) \psi = E \psi,
  \end{align}
  where $V$ is the sum of the external potentials $v^{\rm ext}(\vec{r}_j)$ and electron-electron interactions $v_{\rm ee}(\vec{r}_j,\vec{r}_k)$ of all electrons.
  The marginal wavefunction is then given by
  \begin{align}
    \left( \frac{(-i\pa+\vec{A}(1))^2}{2} + v(1) \right) \chi(1) = E \chi(1)
    \label{eq:chi}
  \end{align}
  which is a one-electron Schr\"odinger equation, where the effect of the other $N-1$ electrons is fully contained in the vector potential
  \begin{align}
    \vec{A}(1) = \Braket{\phi|-i \pa \phi}_{2 \dots N}
  \end{align}
  and the scalar potential $v$, given by
  \begin{align}
    v(1) = v^{\rm H}(1) + v^{\rm G}(1) + v^{\rm ext}(1).
    \label{eq:v}
  \end{align}
  Here,
  \begin{align}
    v^{\rm H}(1) = \Braket{\phi| -\sum_{j=2}^N \frac{\nabla_j^2}{2} + V |\phi}_{2,\dots,N} - v^{\rm ext}(1)
    \label{eq:vh}
  \end{align}
  is the (expectation value of the) energy of $N-1$ electrons given an additional electron is at $\vec{r}_1$, 
  and 
  \begin{align}
    v^{\rm G}(1) = \frac{1}{2} \Braket{\pa \phi | \left( 1 - \ket{\phi} \bra{\phi} \right) | \pa \phi}_{2,\dots,N}
    \label{eq:vg}
  \end{align}
  is connected to the magnitude of change of $\phi(2,\dots,N;1)$ as wavefunction of the $N-1$ electrons at $\vec{r}_2,\dots,\vec{r}_n$ if the position of the additional electron at $\vec{r}_1$ is changed.
  $v^{\rm G}$ is called the geometric potential because it is related to the Fubini-Study metric and the quantum-geometric tensor \cite{provost1980}.
  
  The one-electron potentials in the EEF are obtained from the wavefunction $\phi(2,\dots,N;1)$ of the conditioned system.
  The dependence of this wavefunction on $\vec{r}_1$ reflects the spatial entanglement\cite{schroeder2017} of the conditioned system on the additional electron, i.e., how its state depends on where the additional electron is (measured).
  If the wavefunction of the $N$-electron system was a disentangled state, e.g.\ a Hartree product of one-electron orbitals, $\phi(2,\dots,N;1)$ would not depend on the coordinate $\vec{r}_1$, and $v(1)$ and $\vec{A}(1)$ would be constants; in contrast, a wavefunction for $N$ indistinguishable particles (symmetrized or antisymmetrized) is always entangled, with its conditional wavefunction $\phi(2,\dots,N;1)$ and hence also $v(1)$ and $\vec{A}(1)$ depending on $\vec{r}_1$, at least in some region.
  All features of the potentials reflect this entanglement of one electron with the other electrons.
  
  Importantly, $\chi$ yields the one-electron density,
  \begin{align}
    |\chi(1)|^2 = \rho(1) = \Braket{\psi(1,\dots,N)|\psi(1,\dots,N)}_{2,\dots,N},
  \end{align}
  and (together with $\vec{A}$ and $v^{\rm G}$) the one-electron current density and one-electron observables of the many-electron system.
  The phase of $\chi$ is arbitrary if the phase of $\phi$ is adjusted accordingly (see \eqref{eq:eef}; the magnitude of $\chi$ cannot change because of \eqref{eq:pnc}). 
  This phase is a gauge freedom and we chose the gauge $\vec{A} \stackrel{!}{=} 0$ for simplicity, but we note that this choice is not always possible, e.g., if the system is rotating \cite{requist2016,requist2017}.
  The potentials $v^{\rm H}$, $v^{\rm G}$, and trivially also $v^{\rm ext}$ are gauge-independent.
  
  As the one-electron density of the exact and the KS system are identical by construction, eq.\ \eqref{eq:chi} is (up to a constant shift) also the determining equation of OF-DFT, where $v$ is usually written as
  \begin{align}
    v(1) = v^{\rm P}(1) + v^{\rm HXC}(1) + v^{\rm ext}(1),
    \label{eq:vks}
  \end{align}
  with Hartree-exchange-correlation potential $v^{\rm HXC}(1) = v^{\rm KS}(1) - v^{\rm ext}(1)$.
  As shown below, relation \eqref{eq:vks} is equivalent to \eqref{eq:v}, except that the terms are evaluated for the conditional wavefunction of the KS system.
  The KS potential yields the KS orbitals and KS energies via
  \begin{align}
    \left( -\frac{\pa^2}{2} + v^{\rm KS}(1) \right) \vp_j^{\rm KS}(1) = \ve_j^{\rm KS} \vp_j^{\rm KS}(1).
  \end{align}
  From these orbitals, the wavefunction $\psi^{\rm KS}$ of the non-interacting KS system can be constructed as a Slater determinant.
  As the KS system has, by construction, the same one-electron density like the interacting system,
  \begin{align}
    \rho(1) = \frac{1}{N} \sum\limits_{j=1}^N |\vp_j^{\rm KS}(1)|^2,
    \label{eq:ksdens}
  \end{align}
  we can define the KS conditional wavefunction $\phi^{\rm KS}$ via
  \begin{align}
    \psi^{\rm KS}(1,\dots,N) = \chi(1) \phi^{\rm KS}(2,\dots,N;1).
  \end{align}
  Interpreting $v^{\rm H} = v^{\rm H}[\phi,V]$ and $v^{\rm G} = v^{\rm G}[\phi]$ in \eqref{eq:vh} and \eqref{eq:vg} and thus also $v = v[\phi,V]$ as functionals of $\phi$ and $V$, it follows
  \begin{align}
    v[\phi,V] = v[\phi^{\rm KS},V^{\rm KS}]
    \label{eq:connect}
  \end{align}
  with $V^{\rm KS}(1,\dots,N) = \sum_{j=1}^N v^{\rm KS}(j)$.
  Also, \cite{gritsenko1996b}
  \begin{subequations}
    \begin{align}
      v^{\rm H}[\phi^{\rm KS},V^{\rm KS}] &= v^{\rm PH} + v^{\rm HXC} \label{eq:vhks} \\
      v^{\rm G}[\phi^{\rm KS}]            &= v^{\rm PG}               \label{eq:vgks}
    \end{align}
    \label{eq:vks_rel}
  \end{subequations}
  up to a constant shift, with $v^{\rm PH} + v^{\rm PG} = v^{\rm P}$ and \cite{levy1988,kraisler2020}
  \begin{align}
    v^{\rm PH}(\vec{r}) 
      &= \sum_{n=1}^N (\epsilon_{N}^{\rm KS} - \epsilon_n^{\rm KS}) |\phi_n^{\rm KS}(\vec{r})|^2 
      \label{eq:ph} \\
    v^{\rm PG}(\vec{r}) 
      &= \frac{1}{2} \sum_{n=1}^N |\prv \phi_n^{\rm KS}(\vec{r})|^2,
      \label{eq:pg}
  \end{align}
  where $\phi_n^{\rm KS}(\vec{r}) = \vp_n^{\rm KS}(\vec{r})/\sqrt{\rho(\vec{r})}$.
  Consequently, $v^{\rm PH} + v^{\rm HXC}$ is the (expectation value of the) energy and $v^{\rm PG}$ is the geometric potential of the conditioned KS system.
  
  Relations \eqref{eq:connect} and \eqref{eq:vks_rel} connect the interacting many-electron system to the KS system.
  If $\psi^{\rm KS}$ represents the many-electron problem in a similar way like $\psi$, we expect $v^{\rm H}[\phi,V] \approx v^{\rm H}[\phi^{\rm KS},V^{\rm KS}]$ and $v^{\rm G}[\phi] \approx v^{\rm G}[\phi^{\rm KS}]$; if there is a strong difference between $\psi^{\rm KS}$ and $\psi$, we expect a different partition of $v$ for the interacting and for the KS system.
  
  \begin{figure*}[htb]
    \includegraphics[width=0.99\textwidth]{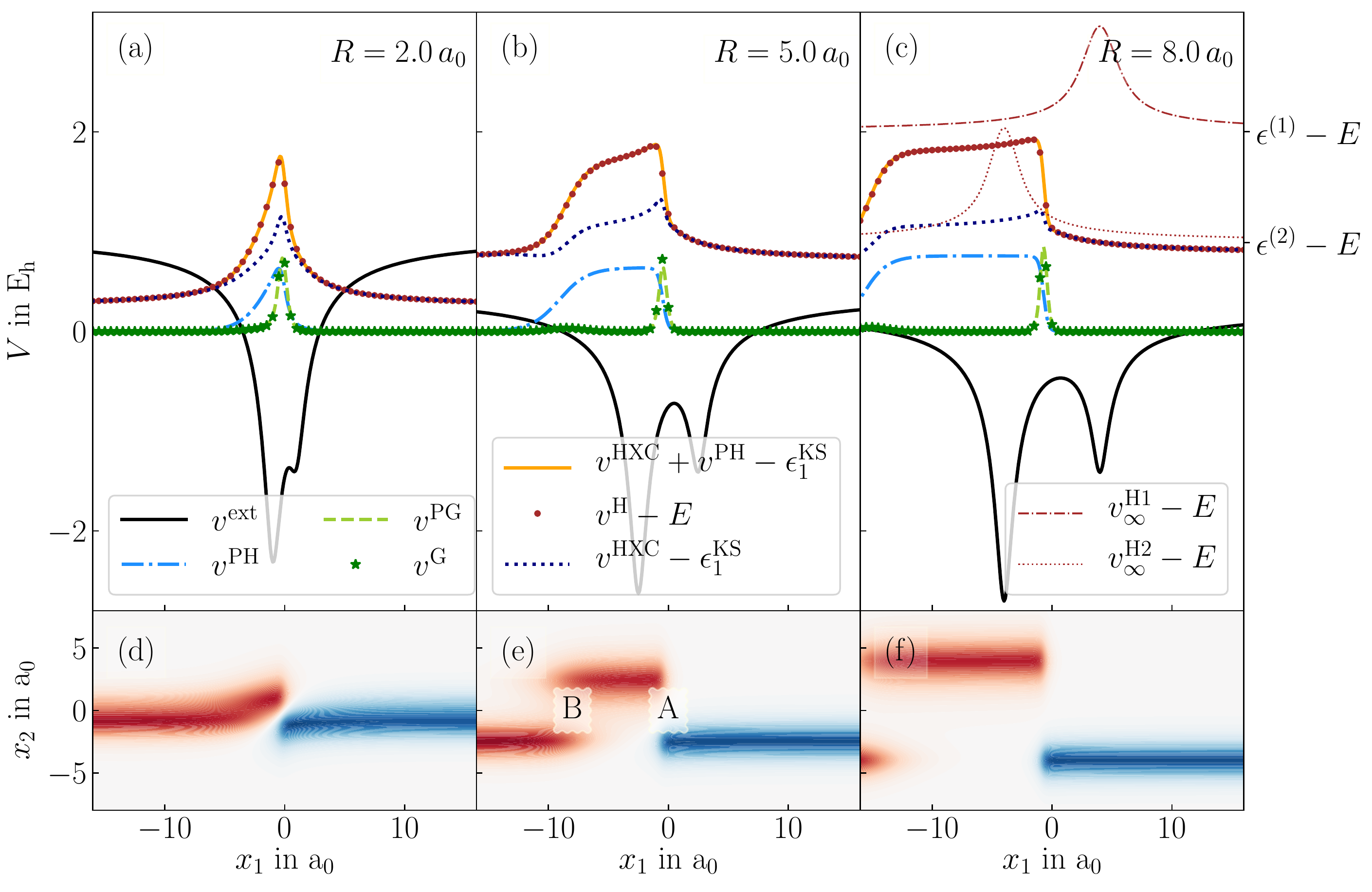}
    \caption{
      Top row: Contributions to the one-electron potential $v(x_1)$ for three values of the internuclear distance $R$ for the diatomic model.
      Bottom row: Conditional wavefunctions $\phi(x_2;x_1)$ of the panels above.
      }
    \label{fig:dafig}
  \end{figure*}
  
  To illustrate how steps form in the potentials during dissociation, we turn to a one-dimensional model of a heteronuclear diatomic molecule with the Hamiltonian
  \begin{align}
    H = \sum_{j=1}^2 \left(-\frac{\partial_j^2}{2} + v_{\rm en}(x_j)  \right) + v_{\rm ee}(x_1,x_2) + v_{\rm nn},
    \label{eq:h_mol}
  \end{align}
  where the electron-nuclear interaction is given by $v_{\rm en}(x) = v_{\rm en}^{(1)}(x,+R/2) + v_{\rm en}^{(2)}(x,-R/2)$ with
  \begin{align}
    v_{\rm en}^{(Z)}(x,R_0) = -\frac{Z}{\sqrt{ (x-R_0)^2 + c_{\rm en} }} ,
    \label{eq:ven}
  \end{align}
  the electron interaction is given by 
  \begin{align}
    v_{\rm ee}(x_1,x_2) = \frac{1}{\sqrt{ (x_1-x_2)^2 + c_{\rm ee} }},
    \label{eq:vee}
  \end{align}
  and the nuclear interaction is given by
  \begin{align}
    v_{\rm nn} = \frac{2}{\sqrt{R^2 + c_{\rm nn}}}.
    \label{eq:vnn}
  \end{align}
  The external potential for this model is $v^{\rm ext}(x) = v_{\rm en}(x) + v_{\rm nn}$ and the parameters are chosen as \unit[$c_{\rm en} = c_{\rm ee} = 0.5$]{$a_0^2$} and \unit[$c_{\rm nn} = 0.1$]{$a_0^2$}.
  The model describes two nuclei with charges $+2$ and $+1$ located at $-R/2$ and $R/2$, respectively, and two electrons at $x_1$ and $x_2$.
  Results for the same model, but with three electrons, can be found in the supplemental material.
  Soft-Coulomb potentials like \eqref{eq:ven}, \eqref{eq:vee}, and \eqref{eq:vnn} are a standard choice when a 3D system is represented with a 1D model \cite{grobe1992}.
  To illustrate the relevant effects, we consider the energetically lowest antisymmetric electronic state.
  
  Panels (a), (b), and (c) of Fig.\ \ref{fig:dafig} show the external potential $v^{\rm ext}$ and the potentials $v^{\rm HXC}$, $v^{\rm PH}$, $v^{\rm PG}$ for the non-interacting KS system as well as the potentials $v^{\rm H}$, $v^{\rm G}$ for the interacting system, for three different values of the internuclear separation $R$.
  We find that $v^{\rm PG} \approx v^{\rm G}$ and $v^{\rm PH} + v^{\rm HXC} \approx v^{\rm H}$ up to a constant, i.e., the interacting system and the KS system describe the one-electron picture in a similar way.
  
  Our focus here is the step structures of the potentials that form when the internuclear distance increases.
  Two steps forming a plateau can be seen in $v^{\rm PH}$ and $v^{\rm HXC}$ as well as in $v^{\rm H}$ for \unit[$R=5$]{$a_0$} and \unit[$R=8$]{$a_0$}.
  One of them is located at $x_1 \approx 0$ (region A in panel (e)) and one is located at smaller $x_1$, e.g.\ for \unit[$R=5$]{$a_0$} at \unit[$x_1 \approx -8$]{$a_0$} (region B in panel (e)).
  The geometric potentials $v^{\rm PG}$ or $v^{\rm G}$ are bell-shaped functions centered at the location of the two steps but are otherwise zero; the bell at $x_1 \approx 0$ is clearly visible, while the one at smaller $x_1$ is barely visible in the figure.
  
  With the KS system, the appearance of the steps in $v^{\rm HXC}$ and $v^{\rm PH}$ can be rationalized based in the asymptotic form of the KS orbitals \cite{hodgson2017,kraisler2020}.
  With the EEF, a mechanistic interpretation of the steps based on $\phi$ can also be given.
  In Fig.\ \ref{fig:dafig} (d), (e), and (f) the conditional wavefunction $\phi(x_2;x_1)$ is shown.
  For \unit[$R=5$]{$a_0$}, for example, $\phi$ shows that the electron at $x_2$ is localized only around \unit[$x_2 = -2.5$]{$a_0$}, only around \unit[$x_2 = +2.5$]{$a_0$}, or localized partially around \unit[$x_2 = -2.5$]{$a_0$} and \unit[$x_2 = +2.5$]{$a_0$} in a wide region around \unit[$x_1\approx-8$]{$a_0$} (region B) and a narrow region around $x_1 \approx 0$ (region A).
  The wavefunction of the conditioned electron at $x_2$ is thus localized on one or both of the two nuclei, depending on where the electron at $x_1$ is found.
  
  \begin{figure}[htbp]
    \includegraphics[width=0.49\textwidth]{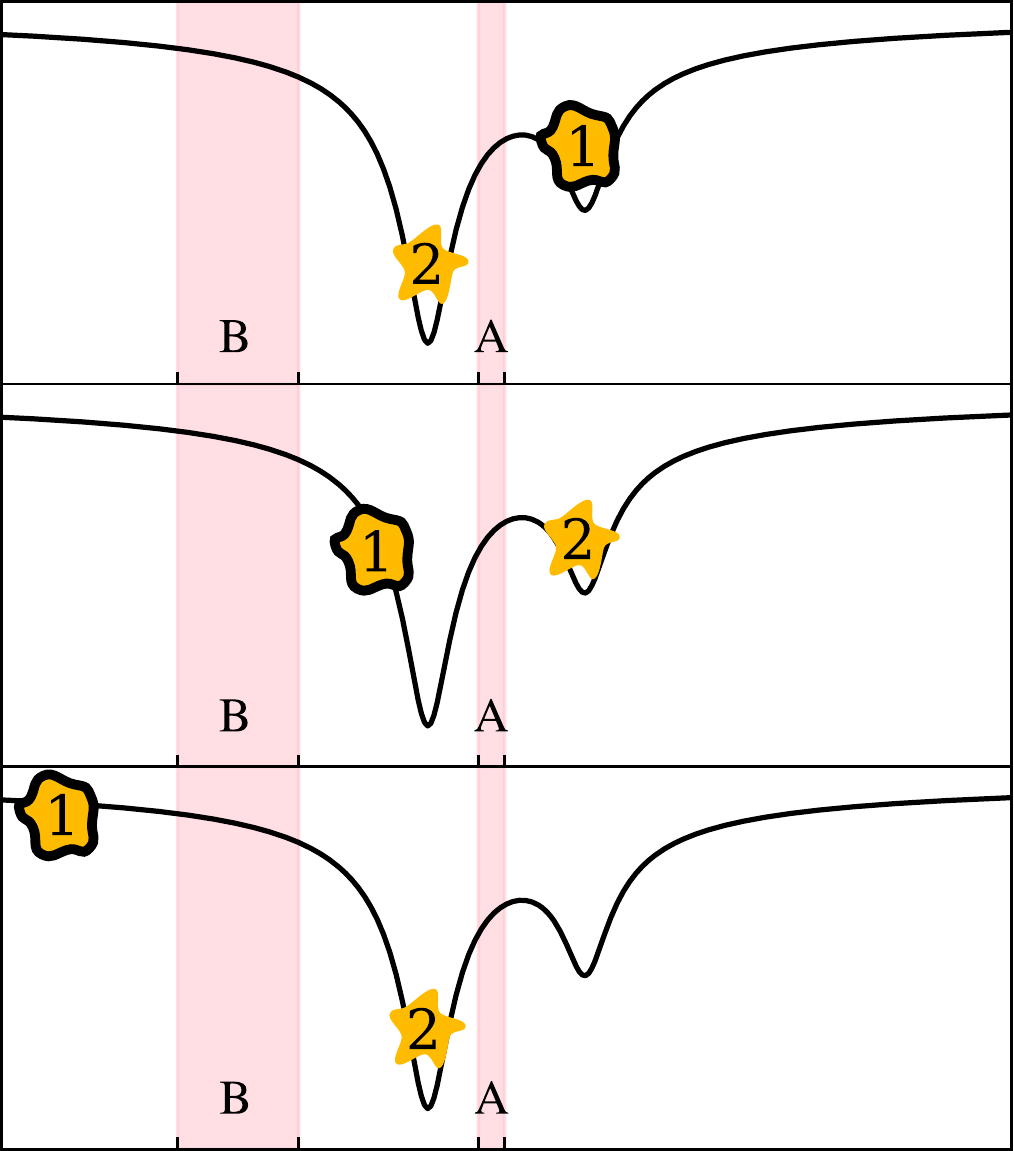}
    \caption{
      Charge-transfer mechanism of the diatomic model.
      The two yellow star shapes represent electrons, where 1 is a condition and 2 is the conditioned system: 
      Given an electron is at a certain place, what is the state of the other electron? 
      The graph is the external potential for the two nuclei with charges $+2$ (left well) and $+1$ (right well).
      If 1 is right of region A, 2 is found in the energetically optimal region around the $+2$-nucleus (top).
      If 1 is between regions B and A, it is close to the $+2$-nucleus and 2 is found in the energetically less favorable region around the $+1$-nucleus (middle).
      If 1 is to the left of region B, 2 is again at the $+2$-nucleus nucleus (bottom).
      The indicated width of the region is $1/\Delta$ with $\Delta = \sqrt{-2 \epsilon_0^{\rm KS}} + c \sqrt{-2 \epsilon_0^{\rm KS}}$, where $c=+1$ for region A and $c=-1$ for region B (see \cite{comppap} for details).
      }
    \label{fig:mech}
  \end{figure}

  From the conditional wavefunction we find the following charge-transfer mechanism for how the step forms for a large internuclear distance (e.g., \unit[$R = 5$]{$a_0$}), illustrated in Fig.\ \ref{fig:mech}:
  When one electron is somewhere between $x_1 = +\infty$ and region A, the conditioned system is in the energetically most favorable configuration where the electron at $x_2$ is localized around the nucleus with charge $+2$ (left well).
  When the electron passes region A and thus comes close to the nucleus with charge +2, the conditioned system changes to the energetically less favorable configuration where the electron at $x_2$ is localized around the nucleus with charge $+1$ (right well).
  Then, when the electron passes region B towards $x_1 = -\infty$, the conditioned system falls back into the energetically more stable configuration.
  If the two electrons were distinguishable by giving them labels according to at which nuclei they are localized, we could also say that the region determines which electron is described with $x_1$ and $x_2$, respectively.
  The sharpness of the step depends on how rapidly the conditional wavefunction $\phi(x_2;x_1)$ changes with $x_1$, with a rapid change in region A and a relatively slow change in region B.\footnote{
    The width of the region can be quantified with the help of the KS orbitals energies, see \cite{comppap}.}
  
  From this mechanism, we can immediately deduce the asymptotic height of the step in $v^{\rm H}$:
  The two steps in $v^{\rm H}$, which appear for larger $R$ and which form a plateau, can be interpreted as the difference between the energy of the conditioned system in the two configurations.
  The plain corresponds to the energy if the conditioned electron is localized at the nucleus with charge $+2$, whereas the plateau corresponds to the energy if the electron is localized at the nucleus with charge $+1$ and thus has a higher energy.
  Thus, for $R \rightarrow \infty$ the step height is (see also \cite{gritsenko1996b})
  \begin{align}
    |\Delta^{\rm H}| = \ve_0^{(1)} - \ve_0^{(2)},
    \label{eq:stepheight}
  \end{align}
  with eigenenergies $\ve_j^{(Z)}$ (indicated on the right ordinate of Fig.\ \ref{fig:dafig} (c)) and eigenstates $\vp_j^{(Z)}$ of the separated atoms centered at the (arbitrary) position $R_0$ given by
  \begin{multline}
    \left(-\frac{\partial_2^2}{2} + v_{\rm en}^{(Z)}(2,R_0) \right) \vp_j^{(Z)}(2;R_0) = \\ \ve_j^{(Z)} \vp_j^{(Z)}(2;R_0).
    \label{eq:vpasy}
  \end{multline}
  Relation \eqref{eq:stepheight} is the sum of the step heights of $v^{\rm KS}$ and $v^{\rm P}$ \cite{hodgson2016,kraisler2020}.
  Those step heights were determined from the analytic asymptotic form of the KS orbitals, while the EEF provides an intuitive picture for the steps.
  
  Based on the charge-transfer mechanism, a way to construct $v^{\rm H}$ for large $R$ suggests itself.
  Using the eigensystem \eqref{eq:vpasy} for the separated atoms, the atomic potentials
  \begin{multline}
    v_{\infty}^{{\rm H}Z}(1;R_0) = \ve_0^{(Z)} \\ + \Braket{\vp_0^{(Z)}(2;R_0)|v_{\rm ee}(1,2)|\vp_0^{(Z)}(2;R_0)}_2,
    \label{eq:asy}
  \end{multline}
  can be constructed. 
  \eqref{eq:asy} is the energy of the atomic one-electron systems with an external electron at $x_1$ if the electron-electron interaction is treated in 1st-order perturbation theory.
  Also, the effect of the kinetic term of the external electron is neglected, which is valid if it is far away from the atom.
  In Fig.\ \ref{fig:dafig} (c), $v_{\infty}^{{\rm H}1}$ and $v_{\infty}^{{\rm H}2}$ are shown with the nuclei placed at the same positions as for the molecule, i.e., $R_0 = R/2$ for $v_{\infty}^{{\rm H}1}$ and $R_0 = -R/2$ for $v_{\infty}^{{\rm H}2}$, respectively.
  To the right and to the left (not shown in the figure) of the plateau $v^{\rm H}$ follows $v_{\infty}^{\rm H2}$ up to a small constant shift, while at the plateau $v^{\rm H}$ follows $v_{\infty}^{\rm H1}$ up to a larger constant shift.
  These energetic shifts vanish for larger $R$.
  
  \begin{figure}[htbp]
    \includegraphics[width=0.5\textwidth]{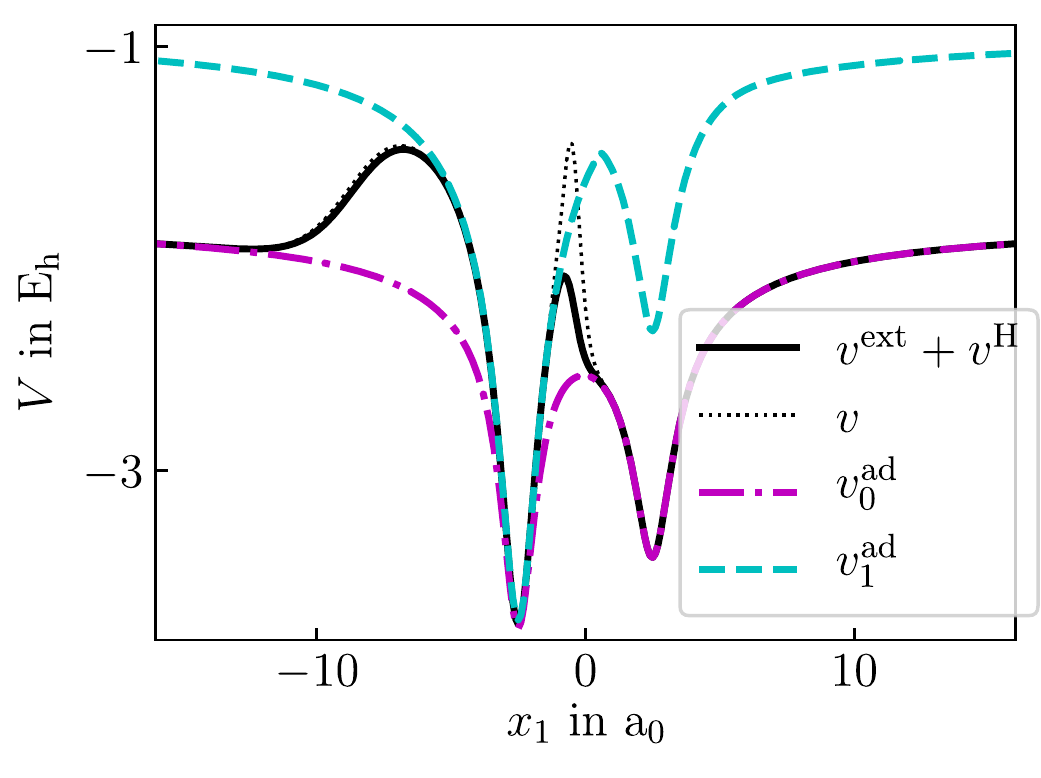}
    \caption{
      Sum of external potential $v^{\rm ext}$ and energy of the conditioned system $v^{\rm H}$, one-electron potential $v$, and lowest two adiabatic one-electron potentials $v_j^{\rm ad}$ for the diatomic with internuclear distance \unit[$R=5$]{$a_0$}.}
    \label{fig:bo}
  \end{figure}
  
  The exact factorization was first used as alternative to the Born-Oppenheimer expansion for the electron-nuclear problem \cite{abedi2010,abedi2012}.
  Based on the formal equivalence of the EEF with the exact electron-nuclear factorization, the EEF also suggests a way to obtain the potential $v$ approximately by using adiabatic states.
  For this purpose, we compute adiabatic states from 
  \begin{align}
    \left(-\frac{\partial_2^2}{2} + V(1,2) \right) \phi_j^{\rm ad}(2;1) = v_j^{\rm ad}(1) \phi_j^{\rm ad}(2;1)
    \label{eq:boa}
  \end{align}
  with $V(1,2) = v_{\rm en}(1) + v_{\rm en}(2) + v_{\rm ee}(1,2) + v_{\rm nn}$ being the 2-electron potential of our model.
  \eqref{eq:boa} is the analogue of the Born-Oppenheimer approximation \cite{born1927} for a system of electrons:
  One electron is clamped and the eigenstates of the other electrons of the system are calculated.
  Fig.\ \ref{fig:bo} shows the lowest two adiabatic potential energy surfaces $v_0^{\rm ad}$, $v_1^{\rm ad}$ together with $v$ and $v^{\rm ext} + v^{\rm H}$ (which is $v$ without the geometric potential $v^{\rm G}$, see \eqref{eq:v}).
  Clearly, $v^{\rm ext} + v^{\rm H}$ is either $v_0^{\rm ad}$ or $v_1^{\rm ad}$, i.e., one of the lowest two adiabatic states, except in the transition regions A and B.
  The steps in $v$ are thus similar to the steps found in the (time-dependent) electron-nuclear problem when a nuclear wavepacket splits and evolves on two adiabatic surfaces in the Born-Oppenheimer representation \cite{abedi2013}.
  Then, the exact potential determining the nuclear state is composed of parts of several adiabatic surfaces in neighboring spatial regions that are connected by steps.
  
  In contrast to the electron-nuclear problem, however, all electrons have the same mass.
  One consequence is that $v^{\rm G}$, which is proportional to the inverse mass(es) of the particles in the marginal system, is non-negligible in the EEF. 
  Due to their heavy mass, the nuclei are little influenced by the rearrangement of the electronic system when the nuclei change position.
  In contrast, in the EEF one electron at $x_1$ feels the rearrangement of the conditioned $(N-1)$-electron system when its position changes, especially if this happens over a small region of $x_1$ like in region A.
  We thus expect that the importance of $v^{\rm G}$ grows with increasing number of electrons $N$.
  
  To summarize, EEF and DFT are connected by the same one-electron potential $v$, but constructed with the interacting or the KS conditioned system, respectively.
  We studied a model of a diatomic and found that steps appearing in $v$ during dissociation can be explained as charge transfer encoded in the conditional wavefunction $\phi$ that represents the spatial entanglement of the electrons.
  This intuitive picture provides both the step height and the approximate behavior of $v$ for large internuclear distances.
  The formal analogy of the EEF with the electron-nuclear problem suggests that $v$ can be reproduced with an adiabatic approximation and that methods developed for the electron-nuclear probem may also be useful for an electrons-only problem.
  Consequently, an expansion in adiabatic states may serve as a computational tool for the construction of the one-electron potentials and may further deepen the understanding of the steps.
  We thus conclude that the connection of the EEF and DFT is very fruitful and should be explored further for the benefit of both theories.
  
  \begin{acknowledgement}
    AS thanks Denis Jelovina (ETH Z\"urich) for helpful discussions.
    This research is supported by an Ambizione grant of the Swiss National Science Foundation (SNF).
  \end{acknowledgement}
  
  \begin{suppinfo}
    Information on the solution of a three-electron model.
  \end{suppinfo}
  
  \bibliography{lit}{}
  \bibliographystyle{unsrt}

\end{document}